\documentclass[manuscript]{aastex63}
\usepackage{amsmath}
\usepackage{multirow}

\begin{document}

% working title -- completely open to changes
\title{Anomalous intensities in the infrared emission of CH$^+$ explained by quantum nuclear motion and electric dipole calculations} 

\author{P. Bryan Changala}
\affiliation{Center for Astrophysics \textbar{} Harvard \& Smithsonian, Cambridge, MA 02138, USA}

\author{David A. Neufeld}
\affiliation{Department of Physics \& Astronomy, Johns Hopkins University, Baltimore, MD 21218, USA}

\author{Benjamin Godard}
\affiliation{Observatoire de Paris, PSL University, CNRS, UMR 8112, F-75014, Paris, France}
\affiliation{Sorbonne Universit\'{e}s, UPMC Univ. Paris 6, UMR 8112, LERMA, F-75005, Paris, France}

%\correspondingauthor{David A. Neufeld}
%\email{neufeld@jhu.edu}
\clearpage
\begin{abstract}

The unusual infrared emission patterns of CH$^+$, recently 
detected in the planetary nebula NGC~7027, are examined 
theoretically with high-accuracy rovibrational wavefunctions 
and \textit{ab initio} dipole moment curves. The calculated 
transition dipole moments quantitatively reproduce the 
observed $J$-dependent intensity variation, which is ascribed 
to underlying centrifugal distortion-induced interference 
effects. We discuss the implications of this anomalous behavior 
for astrochemical modeling of CH$^+$ production {and 
excitation, and provide a simple expression to estimate the 
magnitude of this effect for other light diatomic molecules 
with small dipole derivatives.
\vfill 
\phantom{Extra text to push pagination * Extra text to push pagination * Extra text to push pagination * 
Extra text to push pagination * Extra text to push pagination * Extra text to push pagination * 
Extra text to push pagination * Extra text to push pagination * Extra text to push pagination * }}

\end{abstract}

\vskip 0.5 in

\section{Introduction}
While spectroscopy of pure rotational transitions -- performed at radio and submillimeter wavelengths --  remains the primary tool for studying astrophysical molecules, 
infrared spectroscopy of rovibrational transitions has
proven a valuable additional probe.  Rovibrational transitions of
H$_2$ have long been used to study molecular gas in photodissociation 
regions and interstellar shock waves \citep[e.g.][and references therein]{Geballe2017}; and other molecules studied
at infrared wavelengths include CO \citep[e.g.][]{Pontoppidan2011}, H$_2$O \citep[e.g.][]{Indriolo2015}, HCl \citep{Goto2013}, HF \citep{Indriolo2013}, C$_2$H$_2$, HCN \citep[e.g.][]{Cernicharo1999}, and CO$_2$ \citep[e.g.][]{Sonnentrucker2006} among other molecules. 
In addition to these neutral species, two molecular ions
have been detected recently in the planetary nebula NGC 7027 through observations
of their rovibrational emissions: HeH$^+$ \citep{Neufeld2020} and CH$^+$ 
\citep[][hereafter N21]{Neufeld2021}
For diatomic molecules with a $^1\Sigma$ ground state 
(e.g. CO, HCl, HF, HeH$^+$ and CH$^+$), the various 
transitions within a given vibrational band are conveniently characterized by a 
single index, $m$, defined as $-J_l=-J_u-1$ for $P$-branch transitions and $J_l+1=J_u$
for $R$-branch transitions.  Here, $J_l$ and $J_u$ are the rotational quantum numbers for 
the lower and upper states.

In analysis of astronomical observations, it is often assumed that the relative 
strengths of the various observable transitions within a given vibrational band
are simply governed by the transition wavelengths, $\lambda$, and the 
rotational H\"onl-London factors, $f_{\rm HL} = \vert m \vert$; 
and that the spontaneous radiative rates, $A_{ij}$, are simply
proportional to $\vert m \vert/\lambda^3$.   
If we compare the ratio of a pair of
optically-thin rovibrational line fluxes originating in the 
same upper state, $(v_u,J_u)$,
the dependence assumed above implies that their relative line 
fluxes, $F$, will be proportional to 
$h \nu A_{ij} \propto \vert m \vert /\lambda^4$.  While applicable in many astrophysical
circumstances, this assumption implicitly neglects any dependence of the vibronic
transition dipole moment, $\mu_{\rm vib}$,  on the rotational state of the molecule (i.e.\ on the
index $m$).  

Although the neglect of $m$-dependence in $\mu_{\rm vib}$ is a 
good approximation for many cases of astrophysical relevance, 
studies of a number of diatomic systems~\citep{Herman1955,LeRoy1975,Chackerian1983,Goldman1998,Li2013,Medvedev2017} 
have shown that centrifugal distortion can lead to a significant break-down of this approximation, particularly when $\mu_{\rm vib}$ is small.
In this circumstance, the fractional variation of $\mu_{\rm vib}$ with $m$ can become very large, 
and indeed $\mu_{\rm vib}$ can switch sign at some critical value, $m=m^*$.  
When $\log A_{ij}$
is plotted as a function of $m$, this leads to the ``cusp-like" behavior at $m=m^*$
described by \citet[hereafter M17]{Medvedev2017}.  M17 gave several examples 
of vibrational bands
of DF, HF and CO for which anomalous rovibrational line ratios were expected and/or observed
in laboratory spectra.  While rovibrational emission lines of HF and CO have been detected
from astrophysical sources, the anomalies reported by M17 were of 
little astrophysical relevance.  In the vibrational bands that are strong enough to detect 
astrophysically, the value of $m^*$ was found to be large (e.g. $\sim 35$ for HF $v=1-0$); thus 
the fractional dependence of $\mu_{\rm vib}$ on $m$ was weak for the small values of $m \ll m^*$ 
typically observed in astrophysical spectra.

As reported in the companion paper N21, we have recently observed
large anomalies in the relative strengths of $P$- and $R$-branch line emissions
detected from CH$^+$ molecular ions in NGC 7027; in particular, the $R$-branch lines 
are observed to be significantly weaker than expected under the assumption of an 
$m$-independent $\mu_{\rm vib}$.  Anomalous line ratios are conveniently revealed by the
ratio 

\begin{equation}
\alpha = {(F\lambda^4/\vert m \vert)_R \over (F\lambda^4/\vert m \vert)_P},\label{eq:alpha}
\end{equation}
where the subscripts $R$ and $P$ refer to the $R(J_u-1)$ and $P(J_u+1)$ transitions,
which originate in the same $J=J_u$ upper state.  If $\mu_{\rm vib}$ were independent of $m$,
$\alpha$ would equal unity.

\begin{figure}[htbp]
    \centering 
    \includegraphics[width=5.7in]{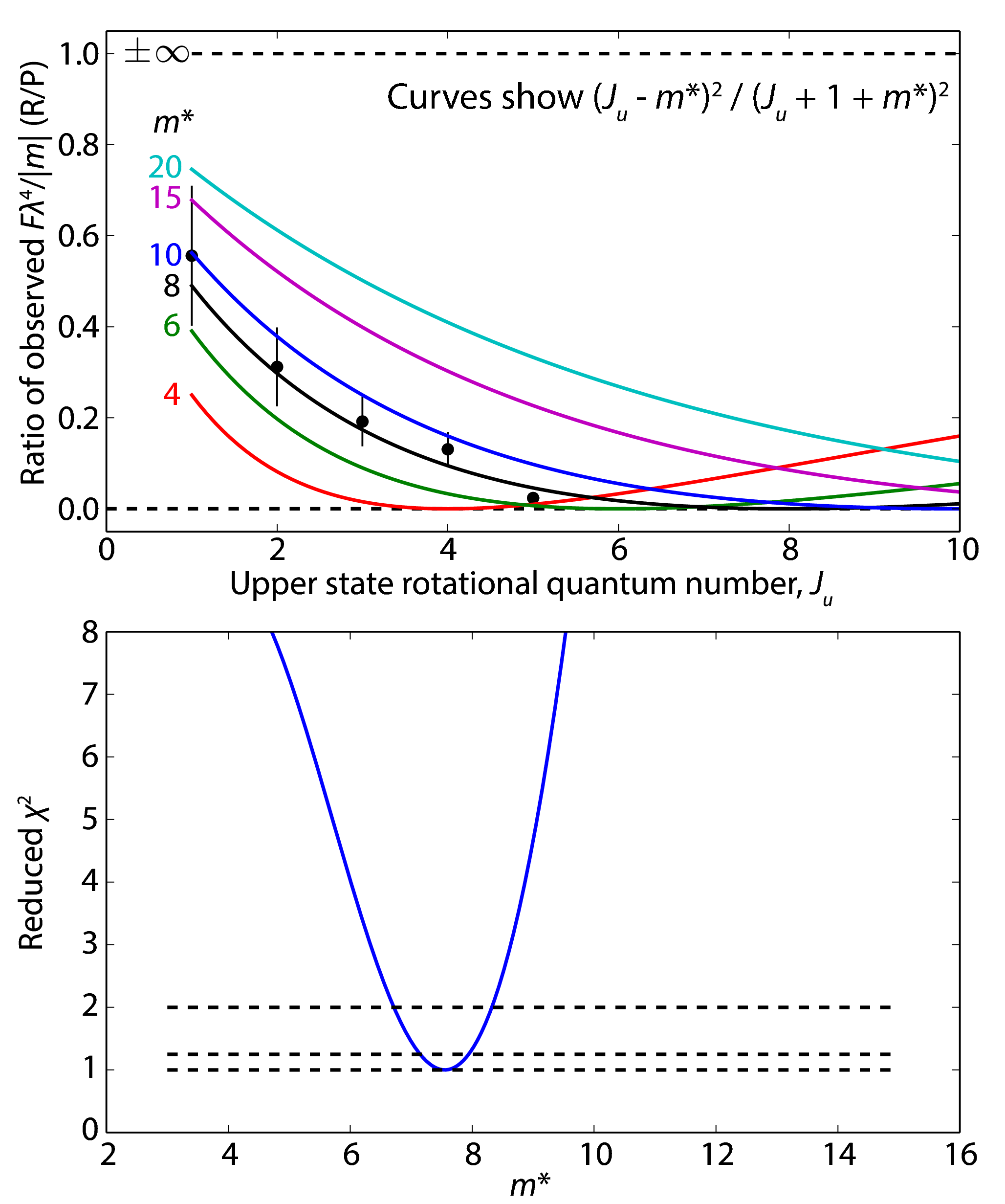}
    \caption{Top panel: Observed $R(J_u-1)/P(J_u+1)$ line ratios, with fluxes scaled by $\lambda^4/\vert m \vert$
    (i.e. the quantity $\alpha$, defined by equation~\ref{eq:alpha}).  Colored curves: Expected behavior for
    several values of $m^*$, under the assumption that the transition dipole moment varies linearly with $m$.
    Lower panel: Reduced $\chi^2$ as a function of $m^*$.  Horizontal dashed lines intersect the curve
    at the best estimate of $m^*$ and at the 68\% and 95\% confidence limits.}
    \label{fig:ratios}
\end{figure}

In Figure~\ref{fig:ratios}, $\alpha$ is plotted for several line pairs observed from
NGC 7027: $R(0)/P(2)$, $R(1)/P(3)$, $R(2)/P(4)$, $R(3)/P(5)$, and $R(4)/P(6)$.
The value of $\alpha$ differs from unity at a high level of statistical significance,
and shows a highly significant variation with $m^*$.
{The Einstein coefficient $A_m$ for a given $R$ or $P$ transition is proportional to $\lambda^{-3} m (2m+1)^{-1} \vert \mu_{\rm vib}(m) \vert^2$ \citep{Hilborn1982}, where $m = m_R = +J_u$ for the $R(J_u-1)$ transition and $m = m_P = -J_u-1$ for the $P(J_u+1)$ transition. If we approximate the $m$-dependence of $\mu_{\rm vib}$ as being linear, $\mu_{\rm vib}(m) \propto (m-m^*)$,} the ratio $\alpha$ simply depends upon the single parameter $m^*$ 
in accordance with 

\begin{align}
\alpha & = {(F\lambda^4/\vert m \vert)_R \over (F\lambda^4/\vert m \vert)_P}\nonumber \\
&= \frac{(m_R/\vert m_R \vert)(2m_R+1)^{-1}\vert \mu_{\rm vib}(m_R) \vert^2}{(m_P/\vert m_P \vert)(2m_P+1)^{-1}\vert \mu_{\rm vib}(m_P) \vert^2}\nonumber \\
&= \frac{+1}{-1}\left(\frac{2J_u + 1}{-2J_u - 1}\right)^{-1}\left\vert\frac{J_u - m^* }{ -J_u -1 - m^* }\right\vert^2 \nonumber \\ 
&=\biggl({J_u-m^* \over J_u+1+m^*}\biggr)^2.\label{eq:alpha_fit}
\end{align}

This dependence is shown by the colored curves in Fig.~\ref{fig:ratios} for different values of $m^*$.  We
have performed a $\chi^2$ analysis to determine the value of $m^*$ suggested by the
astrophysical observations.  Here, we assumed that the uncertainties in the measured
line ratios were the quadrature sum of a statistical and a systematic error.  The latter
was assumed to have a constant fractional value that we determined by requiring a 
minimum reduced-$\chi^2$ of unity.
The fractional systematic uncertainty thereby required, 0.27, is in line with expectations 
for the likely flux calibration uncertainties.
The reduced $\chi^2$ is plotted in the bottom panel of Fig.~\ref{fig:ratios}, and yields a value of
$7.55 \pm 0.8$ (95\% confidence limit) for $m^*$.  
Thus the astrophysical observations alone suggest a rather specific behavior
for $\mu_{\rm vib}$, which appears to switch sign at an unusually small value of $m^*$.
This result motivates the present study.

In this paper, we investigate the anomalous transition intensities of CH$^+$ with a high-level calculation based on numerical rovibrational wavefunctions and \textit{ab initio} dipole moment curves. Details of the transition dipole moment calculations are summarized in Section~\ref{sec:methods}. The results reported in Section~\ref{sec:results} confirm that the observed infrared transition intensity variations (N21) are due to centrifugal distortion-induced interference in the dipole moment matrix elements. In Section~\ref{sec:disc}, we discuss the implications of these results for astrophysical modeling of CH$^+$ and present a simple expression for estimating the magnitude of this effect using spectroscopic parameters routinely computed for diatomic molecules.
\vfill\eject

\section{Methods}
\label{sec:methods}

The Born-Oppenheimer potential energy and dipole moment curves for the $X\,{}^1\Sigma^+$ ground state of CH$^+$ were computed with the coupled cluster method including all-electron single, double, and triple excitations (CCSDT) \citep{Noga1986,Scuseria1988} using Dunning's quadruple-zeta correlation-consistent core-valence basis sets (cc-pCVQZ) \citep{Woon1995, Peterson2002}. Although the focus of this paper is infrared transitions in the $X$ state, we also computed the energy and dipole moment curves for the first excited state, $A\,{}^1\Pi$, as well as the $A-X$ transition dipole moment using the equation-of-motion coupled cluster method at the same level of correlation (EOM-CCSDT) \citep{Bomble2004,Kowalski2001,Kucharski2001}. Each of these curves was evaluated on a grid with respect to the CH$^+$ internuclear distance, $r$, spanning $0.76$ to 2.26~\AA{} in steps of $\Delta r = 0.02$~\AA, with additional points at $r =$ 2.5, 3.0, and 3.5~\AA. The $X$ state equilibrium internuclear distance is $r_e = 1.128$~\AA. All electronic structure calculations were performed with the \textsc{cfour} program package \citep{Matthews2020}. {{\normalfont \textsc{cfour}} input files and output data are available at the external repository of \cite{changala_p_bryan_2021_4725806}.} In addition to the \textit{ab initio} potential energy curves, we employed Cho and Le~Roy's empirical potential curves for the $X$ and $A$ states (hereafter referred to as CLR), which are based on a sophisticated treatment of long-range and non-Born-Oppenheimer effects fitted to high-accuracy spectroscopic data of multiple isotopologues \citep{Cho2016}.

Vibrational wavefunctions $\psi_{i,v,J}(r)$ for a given electronic state $i${, vibrational index $v$,} and total angular momentum quantum number $J$ were calculated by solving for the eigenfunctions of the effective radial Hamiltonian, 
\begin{widetext}
\begin{equation}
\hat{H}_{i,J} = -\frac{\hbar^2}{2 \mu} \frac{\partial^2}{\partial r^2} + V_i(r) + \frac{\hbar^2 }{2 \mu r^2} \left(J(J+1) - \Lambda^2\right)\left(1 + g_i(r)\right),\label{eq:H}
\end{equation}
\end{widetext}
where $\mu$ is the reduced mass (computed with neutral atomic masses) and $\Lambda$ is the projection of the electronic orbital angular momentum on the internuclear axis (i.e., $\Lambda = 0$ in the $^1\Sigma^+$ state and $\Lambda = 1$ in the $^1\Pi$ state). $V_i(r)$ is the adiabatic potential energy surface for the $i^{th}$ electronic state, which may include mass-dependent non-Born-Oppenheimer contributions. Similarly, $g_i(r)$ represents centrifugal non-Born-Oppenheimer corrections. For the $^1\Pi$ state, $V_i(r)$ also contains an effective parity-dependent contribution that reproduces the observed $e$/$f$ $\Lambda$-doubling \citep{Cho2016}. Numerical eigenfunctions of Eq.~\ref{eq:H} were computed with a sinc discrete-variable representation (DVR) basis \citep{Colbert1992} of 250 grid points spanning $r = 0.6$ to 6.0~\AA. Vibronic dipole matrix element integrals $ \mu_\mathrm{vib} = \langle \psi_{i,v,J} \vert \mu_{ii'} (r) \vert \psi_{i',v',J'} \rangle$, where $\mu_{ii'}(r)$ is the purely electronic integral between states $i$ and $i'$, were evaluated under the usual diagonal DVR quadrature approximation \citep{Light2000}, interpolating the \textit{ab initio} dipole grids via 8$^{th}$-order polynomials in $r$.

\section{Results}
\label{sec:results}

\begin{figure}[htbp]
    \centering
    \includegraphics[width = 3.5in]{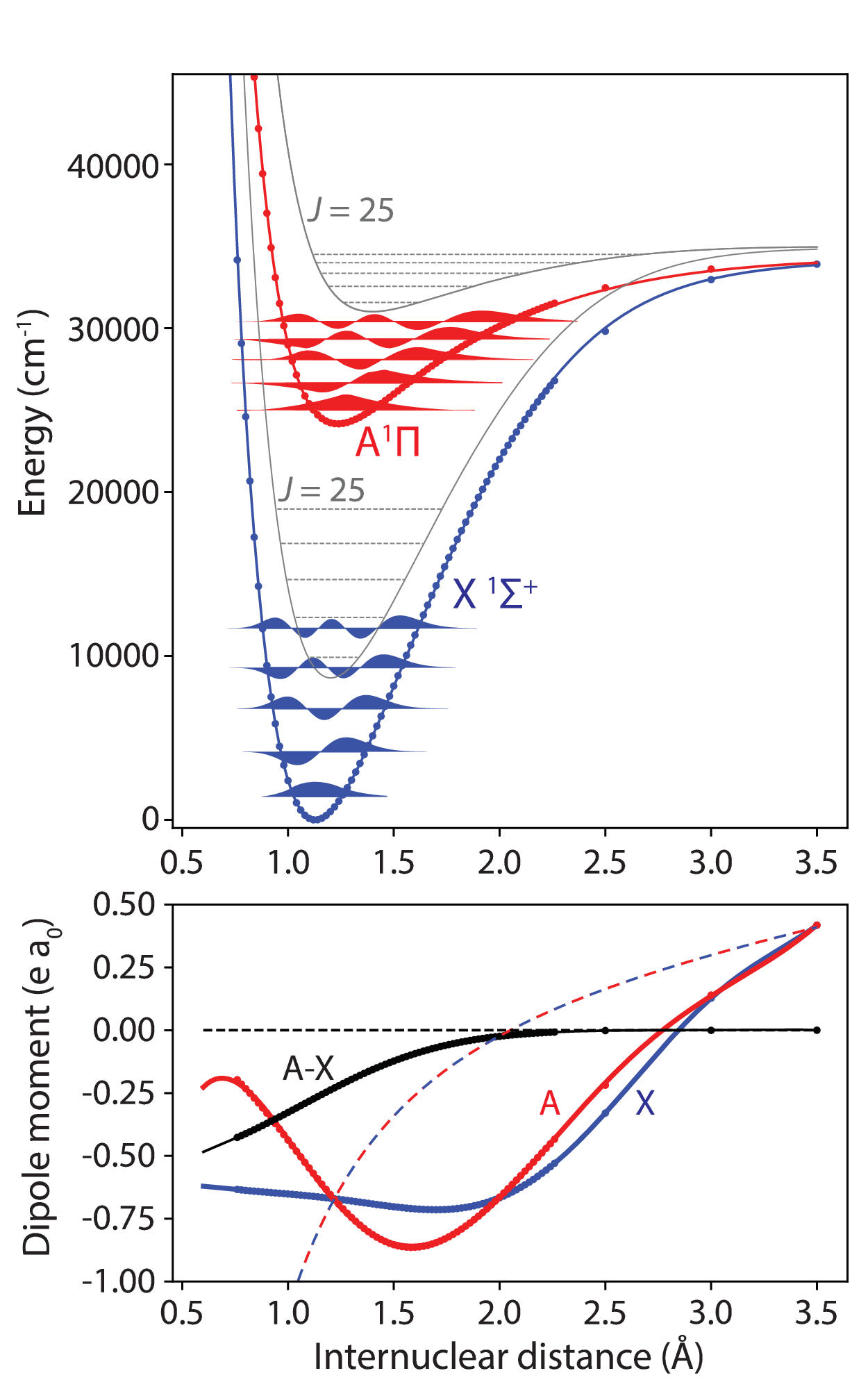}
    \caption{Potential energy and dipole moment curves of CH$^+$. Top panel: The potential energy curves for the ground $X\,{}^1\Sigma^+$ state (blue) and excited $A\,{}^1\Pi$ state (red) are plotted versus internuclear distance along with their lowest few vibrational wavefunctions. \textit{Ab initio} CCSDT and EOM-CCSDT energies are plotted with circles, while the empirical CLR curves are plotted with solid lines. Solid grey curves show the effective potential curves including the centrifugal contribution for $J = 25$. Bottom panel: The $X$ state (blue) and $A$ state (red) permanent dipole moments and the $A-X$ transition dipole (black) are plotted as a function of internuclear distance. Circles represent the \textit{ab initio} values and solid lines are polynomial fits. The dashed lines illustrate the long-range asymptotic behavior of the C$^+$ + H dissociation limit. A positive permanent dipole vector points from H to C. }
    \label{fig:pec}
\end{figure}

Figure~\ref{fig:pec} shows the potential energy curves and the lowest few $J=0$ vibrational wavefunctions for the $X$ and $A$ states. The vibrational energies calculated using the \textit{ab initio} CCSDT electronic energies differ only slightly from those based on the empirical CLR potential for the vibrational states of interest ($\leq 0.1$\% and $\leq 0.3$\% for $v\leq 4$ in the $X$ and $A$ states, respectively). Our final results are based on wavefunctions from the CLR potential.

The permanent and transition dipole curves are also plotted in Fig.~\ref{fig:pec}. Near the equilibrium bond length of $r_{e} \approx 1.13$~\AA, both electronic states have a permanent dipole moment of $\mu \approx -0.6$~$e\,a_0$ (where a negative sign indicates the dipole points toward the H atom), which depends approximately linearly on the bond displacment over $r \approx 0.75 - 1.5$~\AA. At larger internuclear distances, the curves turn over and approach the asymptotic behavior of the C$^+$($^2$P) + H($^2$S) dissociation limit. (The blue-red dashed line in Fig.~\ref{fig:pec} is calculated by summing the dipole of a permanent $+e$ point charge on the C atom with the induced dipole of the H atom having a polarizability volume of $\alpha_{\rm H} = 4.50711$~\AA$^3$ \citep{Schwerdtfeger2019} in the C + H center-of-mass frame.) The $A-X$ transition dipole moment approaches zero in the dissociation limit as it correlates to a $p-p$ transition of C$^+$.

\begin{figure}[h!]
    \centering
    \includegraphics[width = 3.4 in]{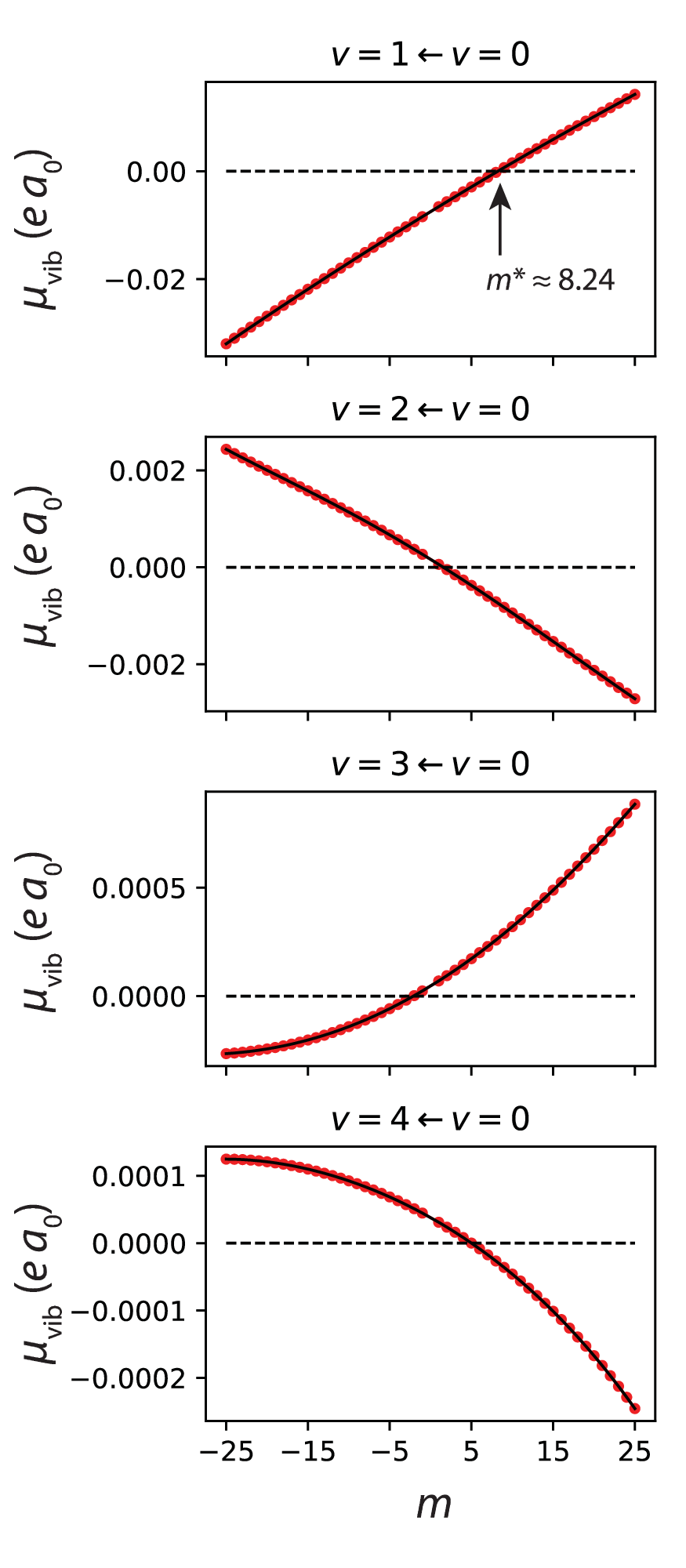}
    \caption{Infrared transition dipole moments for CH$^+$ $X\,{}^1\Sigma^+$. Each plot shows the vibrational transition dipole integrals for R($J$) and P($J$) transitions of the fundamental ($v = 1 \leftarrow 0$) or overtone ($v = 2,3,4 \leftarrow 0$) bands plotted versus their uniform $m$ index. (See text for definition.) }
    \label{fig:muvib}
\end{figure}

Using the calculated vibrational wavefunctions and dipole moment curves, we evaluated the vibronic transition dipole integrals for all rotational, vibrational, and electronic transitions involving states with $v \leq 4$ and $J \leq 25$. Figure~\ref{fig:muvib} shows the vibronic transition dipole moment integrals for the fundamental $v = 1-0$ transition and several overtone transitions within the $X$ state manifold. Each infrared band exhibits a pronounced $m$ dependence. The $v = 1-0$ band, in particular, has an approximately linear dependence on $m$ and crosses zero at $m^* \approx 8.24$. This unusually small value of  $m^*$ is consistent with the astronomically-determined value reported in Section 1.
The dipole integrals for all computed transitions are available in Appendix~\ref{sec:app_sm}.

\section{Discussion}
\label{sec:disc}

\begin{figure}
    \centering 
    \includegraphics[width=5.0in]{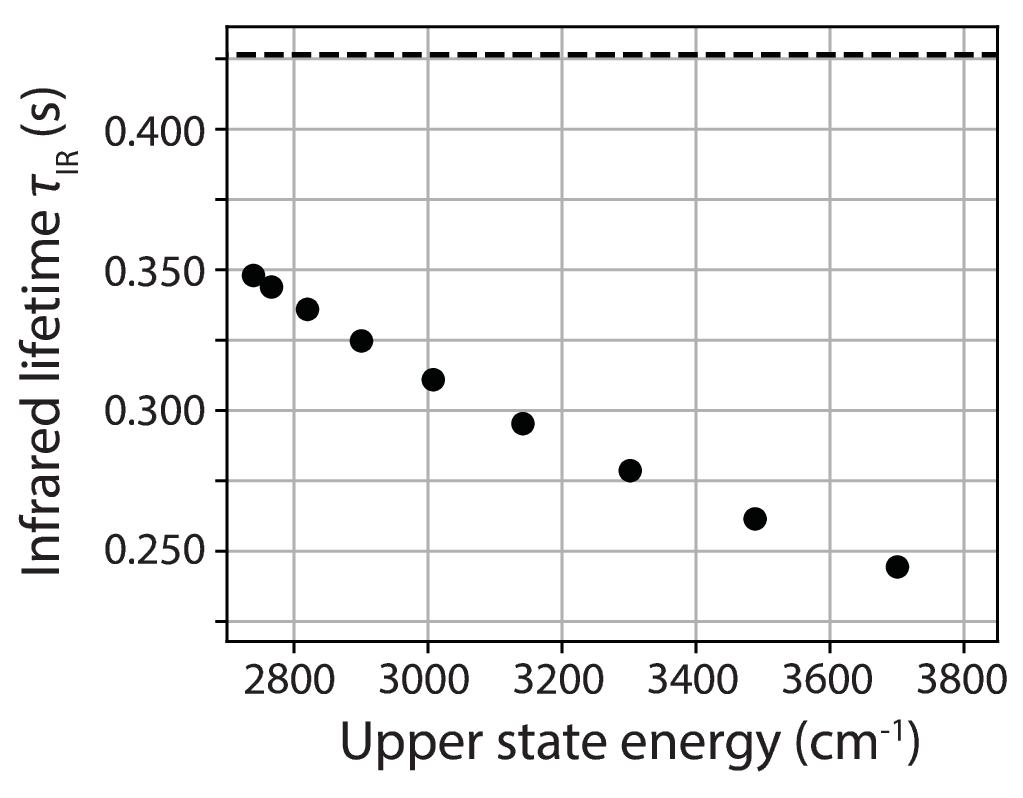}
    \caption{Infrared lifetimes for rotational sub-states of the CH$^+$ $v=1$ level. Black circles are the infrared lifetime accounting for the $m$-dependence of the vibronic transition dipole integral, Eq.~\ref{eq:tauIRJ}. The dashed line is equal to the constant approximation, Eq.~\ref{eq:tauIR0}. Levels with $J_u \leq 8$ are shown.}
    \label{fig:tauIR}
\end{figure}

The calculated transition dipole moment integrals quantitatively reproduce the anomalous intensity patterns of CH$^+$ infrared emission observed in N21. Accurate transition dipole values are essential to estimating molecular column densities from the observed emission flux. They further impact astrochemical modeling of CH$^+$ when spontaneous radiative decay constitutes a significant relaxation pathway. Indeed, the relatively small value of $m^*$ for CH$^+$ has a significant effect on radiative lifetimes. This is illustrated in Fig.~\ref{fig:tauIR} for the $v = 1$ level with the Einstein $A$ coefficient of a given $R$ or $P$ transition calculated as \citep{Hilborn1982}
\begin{equation}
    A_m = \frac{16 \pi^3 \nu_m^3}{3 \epsilon_0 h c^3} \frac{m}{2m+1} \vert \mu_\mathrm{vib}(m) \vert^2,
\end{equation}
where $\nu_m$ is the transition frequency, and the infrared lifetime $\tau_\mathrm{IR}$ is defined as the inverse of the sum of the $A$ coefficients for the $R$ and $P$ transitions sharing a common upper state with angular momentum $J_u$,
\begin{equation}
    \tau_\mathrm{IR} (J_u) \equiv \left( A_{m = J_u} + A_{m = -J_u - 1} \right)^{-1}.\label{eq:tauIRJ}
\end{equation}
A common approximation is to assume that $\vert \mu_\mathrm{vib} \vert$ is $m$-independent and equal to the ``pure vibrational'' value $\mu_{01}$ corresponding to the fictitious $J = 0 - 0$ transition. If one also approximates $\nu_m$ with the fixed band origin $\nu_0$, then the infrared lifetime is constant,
\begin{equation}
    \tau_\mathrm{IR} \approx \left(  \frac{16 \pi^3 \nu_0^3}{3 \epsilon_0 h c^3} \vert \mu_{01} \vert^2 \right)^{-1}.\label{eq:tauIR0}
\end{equation}
This approximation performs poorly for CH$^+$, as Figs.~\ref{fig:muvib} and \ref{fig:tauIR} demonstrate. 

The small moment of inertia of the CH$^+$ rotor leads to a large $J$-dependent centrifugal potential term that shifts the effective equilibrium internuclear distance to larger values for higher $J$ (see Fig.~\ref{fig:pec}).
The infrared transition dipole integrals for $\Delta J = \pm 1$ transitions involve vibrational wavefunctions that experience a differential centrifugal shift, which renders them non-orthogonal and changes the relative contributions of the equilibrium dipole moment $\mu_0 = \mu_X(r_e)$ and its first derivative $\left( \partial \mu_X /\partial r \right ) _0$ to the total vibrational integral~\citep{Medvedev2017}. Interference between these two contributions leads to the observed $m$-dependence of the transition dipole. A semi-classically equivalent picture is that this interference occurs between (i) the pure vibrational oscillating dipole amplitude-modulated by molecular rotation and (ii) the pure rotational oscillating dipole phase-modulated by molecular vibration (which changes the instantaneous moment of inertia and thus the rotation frequency)~\citep{Herman1955}. As shown in Appendix~\ref{sec:mstar}, an approximate value for the zero-crossing point $m^*$ for the $v = 1-0$ transition of a diatomic molecule is given by
\begin{equation}
    m^* \approx \frac{1}{2} \frac{ r_e^3 }{ \rho^2 } \frac{(\partial \mu / \partial r)_0}{\mu_0},\label{eq:mstar}
\end{equation}
where $\rho \equiv \left( \hbar / \mu \omega \right)^{1/2}$ is the characteristic harmonic vibrational length scale. For CH$^+$, this rough estimate yields $m^* \approx 7.52$, in good agreement (better than 10\%) with the high-level calculated value of $m^* = 8.24$. Evaluating Eq.~\ref{eq:mstar} only requires the equilibrium bond length, $r_e$; harmonic frequency, $\omega$; equilibrium dipole moment, $\mu_0$; and dipole moment derivative, $(\partial \mu / \partial r)_0$. All of these quantities must already be calculated for even the simplest harmonic oscillator-linear dipole estimate of infrared transition intensities. They provide a simple, inexpensive estimate of how strongly the $m$-dependent centrifugal effects influence the transition dipole moments and whether a more elaborate calculation such as that reported here is necessary.
\begin{table}

\caption{The $m^*$ estimate, Eq.~\ref{eq:mstar}, evaluated for several diatomic molecules.\label{tab:mstar_estimates}}
\centering
\begin{tabular}{cccccc}\multirow{2}{*}{Molecule}  & $\omega/2\pi c$  & $r_e$  & $\left(\frac{\partial \mu }{ \partial r}\right)_0/\mu_0$ & $m^*$  & $m^*$ \\
& (cm$^{-1}$) & (\AA) &  (\AA$^{-1}$) & (Eq.~\ref{eq:mstar}) & (expt.)\\
\tableline
CH$^+$ & 2860 & 1.128 & 0.13 & 7.5 & 7.6(8)\\
HF & 4139$^{a}$ & 0.917$^{a}$& 0.84$^{b}$ & 38 & 36(1)$^{c}$\\
OH & 3735$^{a}$ & 0.971$^{a}$ &  0.33$^{d}$ & 16 & 14(1)$^{e}$\\
CO & 2170$^{a}$ & 1.128$^{a}$ & $-33.1${}$^{f}$ & $-10^4$ & --\\
\tableline
\multicolumn{6}{l}{$^a$\cite{Herzberg1950}}\\
\multicolumn{6}{l}{$^b$\cite{Buldakov2006}}\\
\multicolumn{6}{l}{$^c$\cite{Medvedev2017}}\\
\multicolumn{6}{l}{$^d$\cite{Nelson1990}}\\
\multicolumn{6}{l}{$^e$\cite{Nelson1989}}\\
\multicolumn{6}{l}{$^f$\cite{Li2015}}
\end{tabular}
\end{table}

{We expect Eq.~\ref{eq:mstar} to be similarly accurate for other light hydrides. As summarized in Table~\ref{tab:mstar_estimates}, the estimate is correct to about 10\% for HF and OH. For a heavy diatomic molecule with a large dipole derivative, this approximation is poor. For example, for CO we calculate an unphysical value of $m^* \approx -10^{4}$, indicating that any $m$-dependence of its vibrational transition dipole will be determined by higher-order rotation-vibration interactions and dipole-derivative effects~\citep{Li2015, Medvedev2017}.}

\vfill\eject
\section{Conclusions}

Successfully interpreting astronomical observations of molecules requires a rigorous understanding of their spectroscopic properties. A simple, but accurate, physical model (Eq.~\ref{eq:mstar}) of CH$^+$ indicates that its anomalous infrared emission is caused by centrifugal distortion-induced interference effects, exaggerated by a small reduced mass and fractional dipole derivative. High-level quantum calculations accounting for these effects quantitatively reproduce the observed emission intensity patterns. The comprehensive tabulation of rovibronic transition dipoles reported here for the CH$^+$ $X-X$ and $A-X$ systems will provide necessary input to astrochemical models of its formation in {astronomical sources}.

\acknowledgments

P.B.C. is supported by NSF grant AST-1908576. We thank Alexandre Faure for fruitful discussions and carefully reviewing this manuscript.

\bibliography{references}

\begin{thebibliography}{}
\expandafter\ifx\csname natexlab\endcsname\relax\def\natexlab#1{#1}\fi
\providecommand{\url}[1]{\href{#1}{#1}}
\providecommand{\dodoi}[1]{doi:~\href{http://doi.org/#1}{\nolinkurl{#1}}}
\providecommand{\doeprint}[1]{\href{http://ascl.net/#1}{\nolinkurl{http://ascl.net/#1}}}
\providecommand{\doarXiv}[1]{\href{https://arxiv.org/abs/#1}{\nolinkurl{https://arxiv.org/abs/#1}}}

\bibitem[{Bomble {et~al.}(2004)Bomble, Sattelmeyer, Stanton, \&
  Gauss}]{Bomble2004}
Bomble, Y.~J., Sattelmeyer, K.~W., Stanton, J.~F., \& Gauss, J. 2004, J. Chem.
  Phys., 121, 5236, \dodoi{10.1063/1.1780159}

\bibitem[{Buldakov {et~al.}(2006)Buldakov, Koryukina, Cherepanov, \&
  Kalugina}]{Buldakov2006}
Buldakov, M.~A., Koryukina, E.~V., Cherepanov, V.~N., \& Kalugina, Y.~N. 2006,
  Russ. Phys. J., 49, 1230, \dodoi{10.1007/s11182-006-0249-8}

\bibitem[{{Cernicharo} {et~al.}(1999){Cernicharo}, {Yamamura},
  {Gonz{\'a}lez-Alfonso}, {de Jong}, {Heras}, {Escribano}, \&
  {Ortigoso}}]{Cernicharo1999}
{Cernicharo}, J., {Yamamura}, I., {Gonz{\'a}lez-Alfonso}, E., {et~al.} 1999,
  \apjl, 526, L41, \dodoi{10.1086/312360}

\bibitem[{Chackerian \& Tipping(1983)}]{Chackerian1983}
Chackerian, C., \& Tipping, R.~H. 1983, J. Mol. Spectrosc., 99, 431,
  \dodoi{10.1016/0022-2852(83)90324-7}

\bibitem[{Changala {et~al.}(2021)Changala, Neufeld, \&
  Godard}]{changala_p_bryan_2021_4725806}
Changala, P.~B., Neufeld, D.~A., \& Godard, B. 2021, {Data for ``Anomalous
  intensities in the infrared emission of CH+ explained by quantum nuclear
  motion and electric dipole calculation''},  Zenodo,
  \dodoi{10.5281/zenodo.4725806}.
\newblock \url{https://doi.org/10.5281/zenodo.4725806}

\bibitem[{Cho \& {Le Roy}(2016)}]{Cho2016}
Cho, Y.~S., \& {Le Roy}, R.~J. 2016, J. Chem. Phys., 144, 024311,
  \dodoi{10.1063/1.4939274}

\bibitem[{Colbert \& Miller(1992)}]{Colbert1992}
Colbert, D.~T., \& Miller, W.~H. 1992, J. Chem. Phys., 96, 1982,
  \dodoi{10.1063/1.462100}

\bibitem[{{Geballe} {et~al.}(2017){Geballe}, {Burton}, \& {Pike}}]{Geballe2017}
{Geballe}, T.~R., {Burton}, M.~G., \& {Pike}, R.~E. 2017, \apj, 837, 83,
  \dodoi{10.3847/1538-4357/aa619e}

\bibitem[{Goldman {et~al.}(1998)Goldman, Schoenfeld, Goorvitch, Chackerian,
  Dothe, M{\'{e}}len, Abrams, \& Selby}]{Goldman1998}
Goldman, A., Schoenfeld, W.~G., Goorvitch, D., {et~al.} 1998, J. Quant.
  Spectrosc. Radiat. Transf., 59, 453, \dodoi{10.1016/S0022-4073(97)00112-X}

\bibitem[{{Goto} {et~al.}(2013){Goto}, {Usuda}, {Geballe}, {Menten},
  {Indriolo}, \& {Neufeld}}]{Goto2013}
{Goto}, M., {Usuda}, T., {Geballe}, T.~R., {et~al.} 2013, \aap, 558, L5,
  \dodoi{10.1051/0004-6361/201322225}

\bibitem[{Herman \& Wallis(1955)}]{Herman1955}
Herman, R., \& Wallis, R.~F. 1955, J. Chem. Phys., 23, 637,
  \dodoi{10.1063/1.1742069}

\bibitem[{Herzberg(1950)}]{Herzberg1950}
Herzberg, G. 1950, {Molecular Spectra and Molecular Structure. Volume I -
  Spectra of Diatomic Molecules}, 2nd edn. (New York: Van Nostrand)

\bibitem[{Hilborn(1982)}]{Hilborn1982}
Hilborn, R.~C. 1982, Am. J. Phys., 50, 982, \dodoi{10.1119/1.12937}

\bibitem[{{Indriolo} {et~al.}(2013){Indriolo}, {Neufeld}, {Seifahrt}, \&
  {Richter}}]{Indriolo2013}
{Indriolo}, N., {Neufeld}, D.~A., {Seifahrt}, A., \& {Richter}, M.~J. 2013,
  \apj, 764, 188, \dodoi{10.1088/0004-637X/764/2/188}

\bibitem[{{Indriolo} {et~al.}(2015){Indriolo}, {Neufeld}, {DeWitt}, {Richter},
  {Boogert}, {Harper}, {Jaffe}, {Kulas}, {McKelvey}, {Ryde}, \&
  {Vacca}}]{Indriolo2015}
{Indriolo}, N., {Neufeld}, D.~A., {DeWitt}, C.~N., {et~al.} 2015, \apjl, 802,
  L14, \dodoi{10.1088/2041-8205/802/2/L14}

\bibitem[{Kowalski \& Piecuch(2001)}]{Kowalski2001}
Kowalski, K., \& Piecuch, P. 2001, Journal of Chemical Physics, 115, 643,
  \dodoi{10.1063/1.1378323}

\bibitem[{Kucharski {et~al.}(2001)Kucharski, W{\l}och, Musia{\l}, \&
  Bartlett}]{Kucharski2001}
Kucharski, S.~A., W{\l}och, M., Musia{\l}, M., \& Bartlett, R.~J. 2001, J.
  Chem. Phys., 115, 8263, \dodoi{10.1063/1.1416173}

\bibitem[{{Le Roy} \& Vrscay(1975)}]{LeRoy1975}
{Le Roy}, R.~J., \& Vrscay, E.~R. 1975, Can. J. Phys., 53, 1560,
  \dodoi{10.1139/p75-198}

\bibitem[{Li {et~al.}(2013)Li, Gordon, {Le Roy}, Hajigeorgiou, Coxon, Bernath,
  \& Rothman}]{Li2013}
Li, G., Gordon, I.~E., {Le Roy}, R.~J., {et~al.} 2013, J. Quant. Spectrosc.
  Radiat. Transf., 121, 78, \dodoi{10.1016/j.jqsrt.2013.02.005}

\bibitem[{Li {et~al.}(2015)Li, Gordon, Rothman, Tan, Hu, Kassi, Campargue, \&
  Medvedev}]{Li2015}
Li, G., Gordon, I.~E., Rothman, L.~S., {et~al.} 2015, Astrophys. J., Suppl.
  Ser., 216, 15, \dodoi{10.1088/0067-0049/216/1/15}

\bibitem[{Light \& Carrington(2000)}]{Light2000}
Light, J.~C., \& Carrington, Jr., T. 2000, Adv. Chem. Phys., 114, 263,
  \dodoi{10.1002/9780470141731.ch4}

\bibitem[{Matthews {et~al.}(2020)Matthews, Cheng, Harding, Lipparini,
  Stopkowicz, Jagau, Szalay, Gauss, \& Stanton}]{Matthews2020}
Matthews, D.~A., Cheng, L., Harding, M.~E., {et~al.} 2020, J. Chem. Phys., 152,
  214108, \dodoi{10.1063/5.0004837}

\bibitem[{Medvedev {et~al.}(2017)Medvedev, Ushakov, Stolyarov, \&
  Gordon}]{Medvedev2017}
Medvedev, E.~S., Ushakov, V.~G., Stolyarov, A.~V., \& Gordon, I.~E. 2017, J.
  Chem. Phys., 147, 164309, \dodoi{10.1063/1.5000717}

\bibitem[{Nelson {et~al.}(1989)Nelson, Schiffman, \& Nesbitt}]{Nelson1989}
Nelson, D.~D., Schiffman, A., \& Nesbitt, D.~J. 1989, J. Chem. Phys., 90, 5455,
  \dodoi{10.1063/1.456451}

\bibitem[{Nelson {et~al.}(1990)Nelson, Schiffman, Nesbitt, Orlando, \&
  Burkholder}]{Nelson1990}
Nelson, D.~D., Schiffman, A., Nesbitt, D.~J., Orlando, J.~J., \& Burkholder,
  J.~B. 1990, J. Chem. Phys., 93, 7003, \dodoi{10.1063/1.459476}

\bibitem[{Neufeld {et~al.}(2021)Neufeld, Godard, Changala, Faure, Geballe,
  G\"{u}sten, Menten, \& Wiesemeyer}]{Neufeld2021}
Neufeld, D.~A., Godard, B., Changala, P.~B., {et~al.} 2021, ApJ, in press

\bibitem[{{Neufeld} {et~al.}(2020){Neufeld}, {Goto}, {Geballe}, {G{\"u}sten},
  {Menten}, \& {Wiesemeyer}}]{Neufeld2020}
{Neufeld}, D.~A., {Goto}, M., {Geballe}, T.~R., {et~al.} 2020, \apj, 894, 37,
  \dodoi{10.3847/1538-4357/ab7191}

\bibitem[{Noga \& Bartlett(1986)}]{Noga1986}
Noga, J., \& Bartlett, R.~J. 1986, J. Chem. Phys., 86, 7041,
  \dodoi{10.1063/1.452353}

\bibitem[{Peterson \& Dunning(2002)}]{Peterson2002}
Peterson, K.~A., \& Dunning, T.~H. 2002, J. Chem. Phys., 117, 10548,
  \dodoi{10.1063/1.1520138}

\bibitem[{{Pontoppidan} {et~al.}(2011){Pontoppidan}, {van Dishoeck}, {Blake},
  {Smith}, {Brown}, {Herczeg}, {Bast}, {Mandell}, {Smette}, {Thi}, {Young},
  {Morris}, {Dent}, \& {K{\"a}ufl}}]{Pontoppidan2011}
{Pontoppidan}, K.~M., {van Dishoeck}, E., {Blake}, G.~A., {et~al.} 2011, The
  Messenger, 143, 32

\bibitem[{Schwerdtfeger \& Nagle(2019)}]{Schwerdtfeger2019}
Schwerdtfeger, P., \& Nagle, J.~K. 2019, Mol. Phys., 117, 1200,
  \dodoi{10.1080/00268976.2018.1535143}

\bibitem[{Scuseria \& Schaefer(1988)}]{Scuseria1988}
Scuseria, G.~E., \& Schaefer, H.~F. 1988, Chem. Phys. Lett., 152, 382,
  \dodoi{10.1016/0009-2614(88)80110-6}

\bibitem[{{Sonnentrucker} {et~al.}(2006){Sonnentrucker},
  {Gonz{\'a}lez-Alfonso}, {Neufeld}, {Bergin}, {Melnick}, {Forrest}, {Pipher},
  \& {Watson}}]{Sonnentrucker2006}
{Sonnentrucker}, P., {Gonz{\'a}lez-Alfonso}, E., {Neufeld}, D.~A., {et~al.}
  2006, \apjl, 650, L71, \dodoi{10.1086/508616}

\bibitem[{Woon \& Dunning(1995)}]{Woon1995}
Woon, D.~E., \& Dunning, T.~H. 1995, The Journal of Chemical Physics, 103,
  4572, \dodoi{10.1063/1.470645}

\end{thebibliography}
\bibliographystyle{aasjournal}

\clearpage 
\appendix 

% Using A/B/C/... appendix prefixes
% for figures and tables
\restartappendixnumbering
\section{Derivation of \lowercase{$m^*$} estimate}
\label{sec:mstar}

For the purposes of deriving a simple estimate of $m^*$, we approximate
the adiabatic potential energy curve as that of a harmonic oscillator 
with frequency $\omega$ and equilibrium bond length $r_e$,
\begin{equation}
    V_0(r) = \frac{1}{2} \mu \omega^2 (r-r_e)^2,
\end{equation}
and expand the centrifugal potential up to first order in the displacement,
\begin{align}
    V_\text{cent}(r) &= \frac{\hbar^2 J(J+1)}{2 \mu} \frac{1}{r^2} \nonumber\\
    &= \frac{\hbar^2 J(J+1)}{2 \mu r_e^2} \left( 1 - 2\frac{r-r_e}{r_e} + \ldots \right).
\end{align}
{The total effective potential $V_0 + V_\text{cent}$ has a minimum displaced by $\delta r (J)= \frac{\hbar^2}{\mu^2 \omega^2 r_e^3} J(J+1)$ up to this order.} For an $R$ or $P$ rovibrational transition, the difference in the effective potential minimum is 
\begin{equation}
    \Delta r(m) = \frac{2\hbar^2}{\mu^2 \omega^2 r_e^3} m,
\end{equation}
where $m$ is the uniform transition index defined in the main text.
It is convenient to introduce the dimensionless displacement coordinate,
\begin{equation}
    x = (r-r_e) / \rho,
\end{equation}
where $\rho \equiv \sqrt{\hbar / \mu \omega}$ is the characteristic harmonic oscillator length scale. The corresponding $m$-dependent displacement is 
\begin{equation}
    \Delta x(m) = 2 \gamma m,
\end{equation}
where $\gamma = (\rho / r_e)^3$.

The overlap integral between the $v = 0$ harmonic oscillator wavefunction and a displaced $v = 1$ wavefunction is
\begin{equation}
    \langle v = 0 \vert v = 1, \Delta x\rangle = -  \frac{\Delta x}{\sqrt{2}}e^{-\Delta x^2/4},
\end{equation}
while the matrix element of the dimensionless displacement coordinate $x$ is
\begin{equation}
    \langle v = 0 \vert x \vert v=1, \Delta x\rangle = \frac{2-\Delta x^2}{2 \sqrt{2}} e^{-\Delta x^2/4}.
\end{equation}

If the permanent dipole moment function is expanded up to first order displacements about the equilibrium bond length,
\begin{equation}
    \mu(r) \approx \mu_0 + \left(\frac{\partial \mu}{\partial r}\right)_0 (r-r_e),
\end{equation}
then the $m$-dependent vibrational transition dipole for a $v = 1-0$ transition is
\begin{align}
    \mu_\text{vib}(m) &= \langle v = 0 \vert \mu(r) \vert v = 1, \Delta x (m) \rangle \nonumber \\
    &= -m \mu_0 \sqrt{2} \gamma e^{-\gamma^2 m^2} + (1-2\gamma^2 m^2) \frac{\rho}{\sqrt{2}} \left( \frac{\partial \mu}{\partial r}\right)_0 e^{-\gamma^2 m^2}.
\end{align}
For CH$^+$, the dimensionless $\gamma$ parameter is about $10^{-3} \ll 1$, so we can approximate $\exp(-\gamma^2 m^2)$ and $(1 - 2 \gamma^2 m^2) $ as $1 + \mathcal{O}(\gamma^2)$, yielding
\begin{equation}
    \mu_\text{vib}(m) \approx \frac{\rho}{\sqrt{2}}  \left( \frac{\partial \mu}{\partial r}\right)_0 - m \sqrt{2} \mu_0 \gamma.
\end{equation}
The first term is the usual harmonic oscillator-linear dipole expression for a $v = 1-0$ transition. The second term represents the contribution from the permanent dipole induced by differential centrifugal distortion, which spoils the orthogonality of the vibrational wavefunctions for $\Delta J = \pm 1$ transitions. These two terms cancel each other at a zero-crossing value of 
\begin{align}
    m^* &\approx \frac{1}{2} \frac{\rho}{\gamma} \frac{(\partial \mu / \partial r)_0}{\mu_0} \nonumber \\
    &= \frac{1}{2} \frac{r_e^3}{\rho^2} \frac{(\partial \mu / \partial r)_0}{\mu_0}.\label{eq:mstarapp}
\end{align}

The $X$ state of CH$^+$ has an \textit{ab initio} harmonic frequency of $2\pi c \times 2860$ cm$^{-1}$ and an equilibrium bond length of 1.128~\AA. The equilibrium dipole moment and dipole derivative are $-0.662$~$e\,a_0$ and $-0.088$~$e\,a_0$ \AA$^{-1}$, respectively. The $m^*$ estimate of Eq.~\ref{eq:mstarapp} is therefore 7.52.

\section{$X-X$ and $A-X$ transition dipole moments}
\label{sec:app_sm}

\begin{figure}[h]
\centering 
\includegraphics[width=6.5 in]{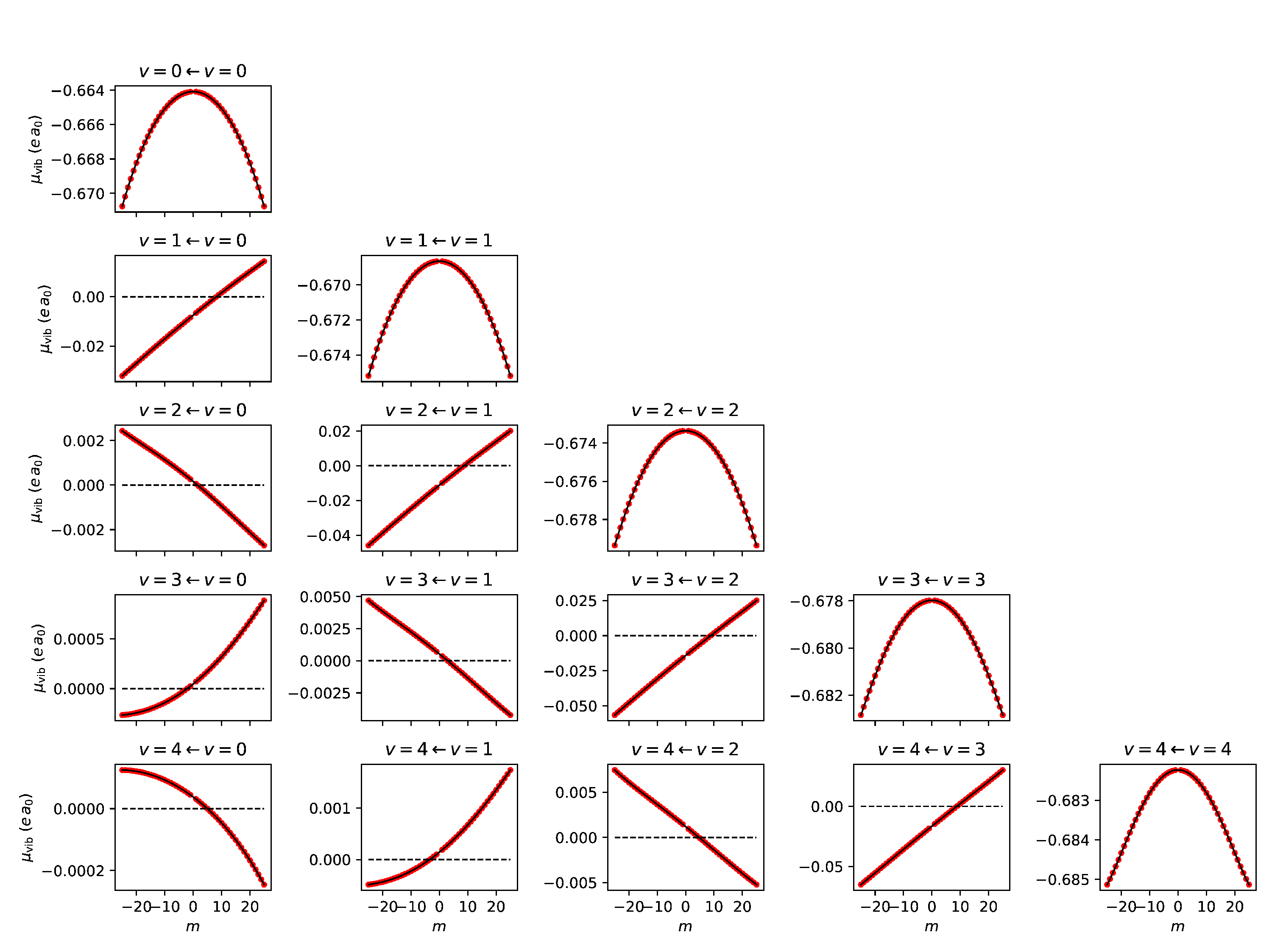}
\caption{\label{fig:suppl_muvib_X} Vibronic transition dipole moments for $X-X$ transitions. The red points are the numerically computed values. The black lines are the result of a 6$^{th}$ order polynomial fit (see Table~\ref{tab:suppl_muvib}).}
\end{figure}

\begin{figure}[h]
\centering 
\includegraphics[width=6.5 in]{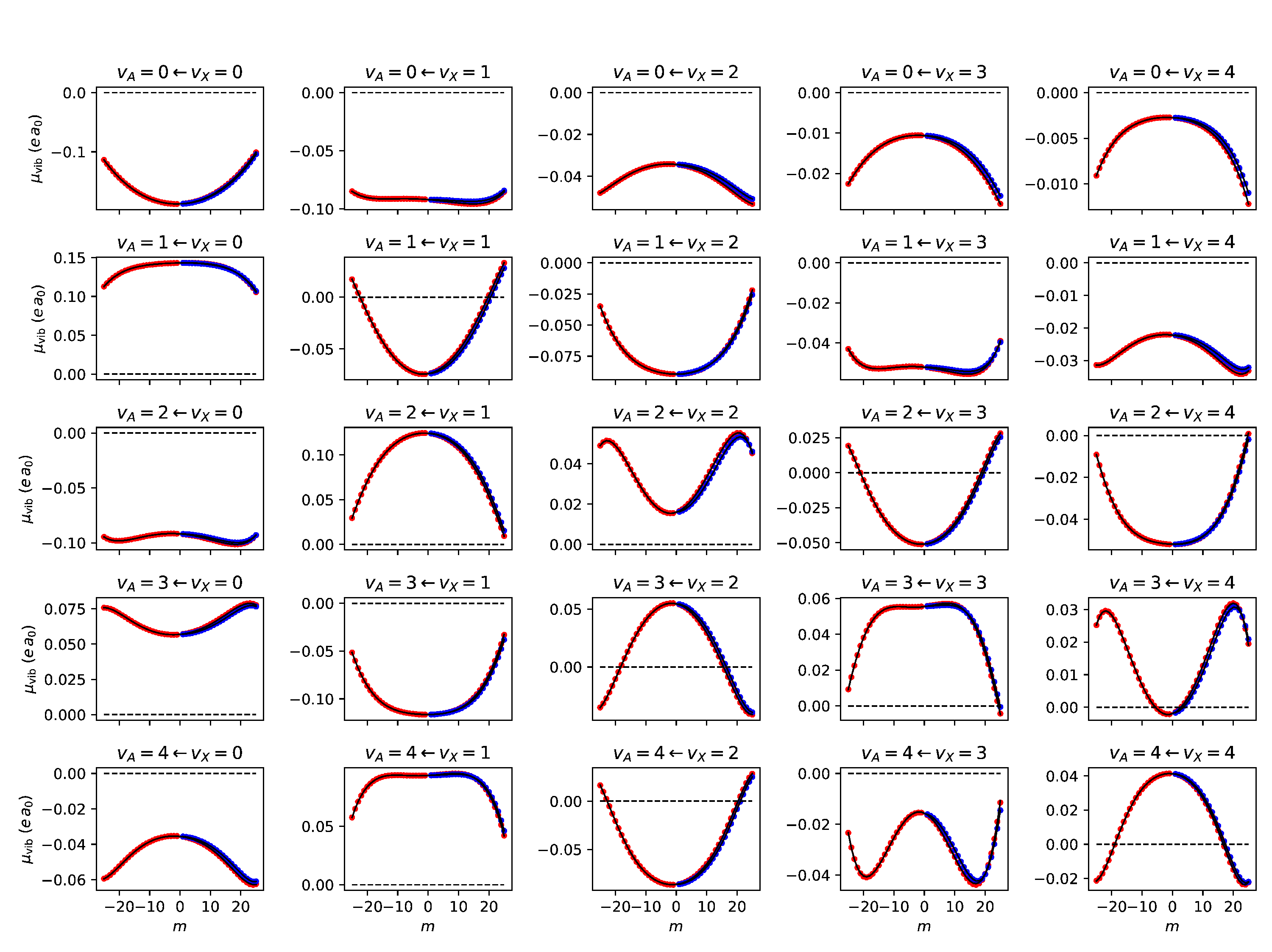}
\caption{\label{fig:suppl_muvib_AX} Vibronic transition dipole moments for $A-X$ transitions. The red points are the numerically computed values for $e-e$ transitions ($R$ and $P$). The blue points are values for $f-e$ transitions ($Q$). The black lines are the result of 6$^{th}$ order polynomial fits (see Table~\ref{tab:suppl_muvib}).}
\end{figure}

This section provides a more complete overview of the calculated transition dipole moments presented in the main text. The vibronic transition dipole moment integrals for pure rotational and rovibrational transitions within the $X$ state manifold for all vibrational quantum numbers $v \leq 4 $ and rotational quantum numbers $J \leq 25$  are plotted in Fig.~\ref{fig:suppl_muvib_X} versus the rotational $m$ index. Those for rovibronic transitions between the $A$ and $X$ states are shown in Fig.~\ref{fig:suppl_muvib_AX}. $e$-parity components of the $A$ state are accessed via $R$ and $P$ transitions from the $X$ state, and we use the standard rotational $m$ index as defined in the main text. For $Q$-branch transitions to the $f$-parity component, we define the rotational index as $m = J_l = J_u$. 

The $m$-dependent transition dipole moments $\mu_\text{vib}(m)$ for each vibronic band were fitted to a 6$^{th}$ order polynomial in $m$. The coefficients of these fits are reported in Table~\ref{tab:suppl_muvib}. The derived Einstein $A$ coefficients and lifetimes for the $v = 1-0$ infrared transitions are summarized in Table~\ref{tab:suppl_Acoeff}. {Plaintext versions of Tables~\ref{tab:suppl_muvib} and \ref{tab:suppl_Acoeff} are available in the data repository~\citep{changala_p_bryan_2021_4725806}.}

% include the CH+ transition dipole moment 
% coefficients table
\begin{table}[h]
\centering
\caption{\label{tab:suppl_muvib} CH$^+$ vibronic transition dipole moments.
          $E'$ ($E''$) is the upper (lower) electronic state label: $0 = X$, $1 = A(e)$, $2 = A(f)$.
          $v'$ ($v''$) is the upper (lower) vibrational quantum number.
          The transition dipoles (in $e\cdot a_0$) for a given $(E',v') \leftarrow (E'',v'')$ vibronic transition
          are represented as a polynomial in the rotational $m$ parameter,
          $\mu_\text{vib} = c_0 + c_1 m + c_2 m^2 + c_3 m^3 + \ldots$.}
\fontsize{6.0}{7.0}\selectfont
% LaTeX table for CH+ transition dipole moments
% [This table was generated with JMAX = 25 on Jan-21-2021 -- PBC]
% 

\begin{tabular}{c@{\,}c@{\,}c@{\,}cc@{\quad}c@{\quad}c@{\quad}c@{\quad}c@{\quad}c@{\quad}c}
 $E'$ & $v'$ & $E''$ & $v''$ & $c_0 $ & $c_1 $ & $c_2 $ & $c_3 $ & $c_4 $ & $c_5 $ & $c_6 $ \\
\hline
  0 &   0 &   0 &   0 & $-6.6408148(-01)$ & $-5.8849997(-18)$ & $-9.8388086(-06)$ & $+3.9539806(-20)$ & $-1.2917802(-09)$ & $-2.8794057(-23)$ & $-7.0560829(-14)$ \\
  0 &   1 &   0 &   0 & $-7.5019291(-03)$ & $+9.2930990(-04)$ & $-2.1290736(-06)$ & $-1.8070449(-08)$ & $-2.1949350(-10)$ & $+1.8575513(-11)$ & $+1.7226015(-13)$ \\
  0 &   2 &   0 &   0 & $+1.6379481(-04)$ & $-1.0441255(-04)$ & $-7.0268509(-07)$ & $+4.4022731(-09)$ & $+3.2595661(-10)$ & $-3.2473527(-12)$ & $+3.4473399(-14)$ \\
  0 &   3 &   0 &   0 & $+4.6435669(-05)$ & $+2.3049349(-05)$ & $+4.3265411(-07)$ & $-3.7627468(-11)$ & $-1.2622394(-11)$ & $-5.3795745(-14)$ & $-8.8052426(-15)$ \\
  0 &   4 &   0 &   0 & $+3.7971100(-05)$ & $-6.8303268(-06)$ & $-1.4497408(-07)$ & $-9.0849348(-10)$ & $-1.8619286(-11)$ & $-2.1778520(-14)$ & $-2.0846905(-15)$ \\
  0 &   1 &   0 &   1 & $-6.6864769(-01)$ & $-6.3371869(-19)$ & $-9.7984370(-06)$ & $-5.4132881(-21)$ & $-1.1971651(-09)$ & $+1.9196038(-23)$ & $+1.8008673(-13)$ \\
  0 &   2 &   0 &   1 & $-1.1196941(-02)$ & $+1.3224555(-03)$ & $-2.9050517(-06)$ & $-1.9811882(-08)$ & $+1.8183465(-11)$ & $+2.6884844(-11)$ & $+3.4150944(-13)$ \\
  0 &   3 &   0 &   1 & $+5.3055497(-04)$ & $-1.7935605(-04)$ & $-8.8572671(-07)$ & $+5.9366336(-09)$ & $+6.2247771(-10)$ & $-6.7397393(-12)$ & $+5.7763473(-14)$ \\
  0 &   4 &   0 &   1 & $+1.4433928(-04)$ & $+4.4846972(-05)$ & $+8.1200137(-07)$ & $-5.9621073(-10)$ & $-5.1790046(-11)$ & $-5.4597844(-14)$ & $-2.5149743(-14)$ \\
  0 &   2 &   0 &   2 & $-6.7335387(-01)$ & $-4.0312123(-18)$ & $-9.2932933(-06)$ & $+1.4198980(-20)$ & $-8.3223898(-10)$ & $+1.9196038(-23)$ & $+5.3240752(-13)$ \\
  0 &   3 &   0 &   2 & $-1.4174558(-02)$ & $+1.6313171(-03)$ & $-3.1236221(-06)$ & $-1.7099133(-08)$ & $+5.2612468(-10)$ & $+3.2130775(-11)$ & $+5.6547438(-13)$ \\
  0 &   4 &   0 &   2 & $+1.1984689(-03)$ & $-2.5256916(-04)$ & $-7.6434913(-07)$ & $+5.3660860(-09)$ & $+9.5175187(-10)$ & $-1.1242882(-11)$ & $+7.3947571(-14)$ \\
  0 &   3 &   0 &   3 & $-6.7797935(-01)$ & $+1.0657235(-18)$ & $-8.0966610(-06)$ & $-9.7335616(-22)$ & $-1.0658901(-10)$ & $+1.9196038(-23)$ & $+1.0004366(-12)$ \\
  0 &   4 &   0 &   3 & $-1.6496603(-02)$ & $+1.8983507(-03)$ & $-2.6888703(-06)$ & $-1.2255954(-08)$ & $+1.3233801(-09)$ & $+3.4623071(-11)$ & $+8.4998223(-13)$ \\
  0 &   4 &   0 &   4 & $-6.8223250(-01)$ & $-8.9873583(-19)$ & $-5.9369154(-06)$ & $-1.2420771(-21)$ & $+1.0714842(-09)$ & $+4.7990095(-23)$ & $+1.5953261(-12)$ \\
  1 &   0 &   0 &   0 & $-1.8806240(-01)$ & $+2.3506381(-04)$ & $+1.1336251(-04)$ & $+7.1318652(-08)$ & $+2.4680124(-08)$ & $-6.2451149(-11)$ & $+1.4761846(-12)$ \\
  1 &   1 &   0 &   0 & $+1.4311079(-01)$ & $+7.6692880(-05)$ & $-1.5009617(-05)$ & $-3.5356689(-07)$ & $-4.7865294(-08)$ & $+5.2209856(-12)$ & $-2.4259898(-11)$ \\
  1 &   2 &   0 &   0 & $-9.1864191(-02)$ & $-2.0712492(-04)$ & $-3.6061494(-05)$ & $+2.5986124(-07)$ & $+2.2188538(-08)$ & $+1.9275226(-10)$ & $+4.8720095(-11)$ \\
  1 &   3 &   0 &   0 & $+5.6861595(-02)$ & $+2.1919237(-04)$ & $+4.9474524(-05)$ & $-6.0456055(-08)$ & $+1.2766953(-08)$ & $-3.6646115(-10)$ & $-6.5442479(-11)$ \\
  1 &   4 &   0 &   0 & $-3.5565385(-02)$ & $-1.8606216(-04)$ & $-4.5712424(-05)$ & $-1.2811552(-07)$ & $-4.3481717(-08)$ & $+5.1800858(-10)$ & $+8.2158366(-11)$ \\
  1 &   0 &   0 &   1 & $-9.2070172(-02)$ & $-2.1221611(-04)$ & $-1.3483612(-05)$ & $+2.7108623(-07)$ & $+2.3130427(-08)$ & $+8.4192083(-11)$ & $+2.5656422(-11)$ \\
  1 &   1 &   0 &   1 & $-7.4103096(-02)$ & $+5.2983588(-04)$ & $+1.6940982(-04)$ & $-6.5104965(-08)$ & $+2.4130323(-08)$ & $-4.4010662(-10)$ & $-6.5424129(-11)$ \\
  1 &   2 &   0 &   1 & $+1.2418851(-01)$ & $-2.8154186(-04)$ & $-1.1744612(-04)$ & $-6.2198278(-07)$ & $-1.2035033(-07)$ & $+6.7589208(-10)$ & $+6.3581538(-11)$ \\
  1 &   3 &   0 &   1 & $-1.1636570(-01)$ & $+8.8894329(-06)$ & $+3.4278280(-05)$ & $+8.9973462(-07)$ & $+1.4379514(-07)$ & $-4.9746990(-10)$ & $-1.3631414(-11)$ \\
  1 &   4 &   0 &   1 & $+9.3439992(-02)$ & $+1.5070068(-04)$ & $+2.1288606(-05)$ & $-8.2049701(-07)$ & $-1.1119732(-07)$ & $+1.1685191(-10)$ & $-5.5751636(-11)$ \\
  1 &   0 &   0 &   2 & $-3.4342416(-02)$ & $-1.5028070(-04)$ & $-2.8455016(-05)$ & $-4.5589799(-08)$ & $-1.0265057(-08)$ & $+1.8651990(-10)$ & $+2.2732055(-11)$ \\
  1 &   1 &   0 &   2 & $-8.9533909(-02)$ & $-1.4461967(-05)$ & $+4.7806030(-05)$ & $+6.2952016(-07)$ & $+8.6428763(-08)$ & $-3.0588407(-10)$ & $-1.0414764(-11)$ \\
  1 &   2 &   0 &   2 & $+1.5906312(-02)$ & $+5.1936140(-04)$ & $+1.4213362(-04)$ & $-8.2346211(-07)$ & $-9.5057502(-08)$ & $-2.1243664(-10)$ & $-8.3856393(-11)$ \\
  1 &   3 &   0 &   2 & $+5.4563562(-02)$ & $-5.7080112(-04)$ & $-1.9261509(-04)$ & $+2.6956314(-08)$ & $-4.8292877(-08)$ & $+1.1102125(-09)$ & $+1.9115645(-10)$ \\
  1 &   4 &   0 &   2 & $-8.6207159(-02)$ & $+3.8245738(-04)$ & $+1.4780180(-04)$ & $+8.5893850(-07)$ & $+1.9430434(-07)$ & $-1.7360651(-09)$ & $-2.4308759(-10)$ \\
  1 &   0 &   0 &   3 & $-1.0621868(-02)$ & $-6.8808516(-05)$ & $-1.6046086(-05)$ & $-9.2339427(-08)$ & $-1.4452656(-08)$ & $+6.8909980(-11)$ & $+5.1576896(-12)$ \\
  1 &   1 &   0 &   3 & $-5.2050924(-02)$ & $-1.5109275(-04)$ & $-1.8206066(-05)$ & $+2.1773079(-07)$ & $+2.7953010(-08)$ & $+2.4504349(-10)$ & $+4.7283955(-11)$ \\
  1 &   2 &   0 &   3 & $-5.1002121(-02)$ & $+2.3348153(-04)$ & $+1.1266508(-04)$ & $+5.5481151(-07)$ & $+9.5768164(-08)$ & $-1.0239612(-09)$ & $-1.3517945(-10)$ \\
  1 &   3 &   0 &   3 & $+5.5466040(-02)$ & $+2.1725136(-04)$ & $+3.0978559(-05)$ & $-1.5971512(-06)$ & $-2.6244429(-07)$ & $+1.2838568(-09)$ & $+1.2203744(-10)$ \\
  1 &   4 &   0 &   3 & $-1.5561693(-02)$ & $-5.4003029(-04)$ & $-1.6130565(-04)$ & $+1.6278459(-06)$ & $+2.4355226(-07)$ & $-5.7097695(-10)$ & $+1.7551911(-11)$ \\
  1 &   0 &   0 &   4 & $-2.7385534(-03)$ & $-2.3655803(-05)$ & $-6.3250990(-06)$ & $-4.9781505(-08)$ & $-7.1028574(-09)$ & $-1.9155816(-11)$ & $-4.8848425(-12)$ \\
  1 &   1 &   0 &   4 & $-2.2186790(-02)$ & $-1.1293752(-04)$ & $-2.4036998(-05)$ & $-8.3581741(-08)$ & $-1.6146978(-08)$ & $+3.2896226(-10)$ & $+4.5871322(-11)$ \\
  1 &   2 &   0 &   4 & $-5.1951276(-02)$ & $-4.0810629(-05)$ & $+2.3120104(-05)$ & $+6.7656177(-07)$ & $+1.1157180(-07)$ & $-4.5157392(-10)$ & $-4.1139299(-11)$ \\
  1 &   3 &   0 &   4 & $-2.0275692(-03)$ & $+3.5196171(-04)$ & $+1.2688894(-04)$ & $-4.5900581(-07)$ & $-7.2402516(-08)$ & $-4.8433730(-10)$ & $-1.1006282(-10)$ \\
  1 &   4 &   0 &   4 & $+4.1290932(-02)$ & $-1.9328688(-04)$ & $-1.0071775(-04)$ & $-9.1085620(-07)$ & $-1.8739555(-07)$ & $+1.9166092(-09)$ & $+2.9888901(-10)$ \\
  2 &   0 &   0 &   0 & $-1.8807202(-01)$ & $+1.1956294(-04)$ & $+1.0955718(-04)$ & $+5.6007270(-07)$ & $-2.0887730(-08)$ & $+1.8799294(-09)$ & $-2.6430667(-11)$ \\
  2 &   1 &   0 &   0 & $+1.4312038(-01)$ & $-2.5571950(-05)$ & $-1.0347307(-05)$ & $-8.2607293(-07)$ & $+1.2035959(-08)$ & $-2.2708029(-09)$ & $+5.7723668(-12)$ \\
  2 &   2 &   0 &   0 & $-9.1848678(-02)$ & $-3.9533481(-05)$ & $-3.4937180(-05)$ & $-2.9290894(-07)$ & $+6.6786147(-08)$ & $-2.2367621(-09)$ & $+9.0386573(-11)$ \\
  2 &   3 &   0 &   0 & $+5.6790510(-02)$ & $+9.3966754(-05)$ & $+3.2860933(-05)$ & $+3.2566227(-06)$ & $-2.9563561(-07)$ & $+1.3093060(-08)$ & $-2.7177335(-10)$ \\
  2 &   4 &   0 &   0 & $-3.5383374(-02)$ & $-1.8614651(-04)$ & $+6.8562614(-06)$ & $-9.8255368(-06)$ & $+8.3036035(-07)$ & $-3.5613336(-08)$ & $+6.2148744(-10)$ \\
  2 &   0 &   0 &   1 & $-9.2056431(-02)$ & $-2.0094685(-05)$ & $-1.0934669(-05)$ & $-4.7145210(-07)$ & $+7.5593205(-08)$ & $-2.3463776(-09)$ & $+6.4842142(-11)$ \\
  2 &   1 &   0 &   1 & $-7.4171017(-02)$ & $+2.0705876(-04)$ & $+1.5369369(-04)$ & $+2.9488146(-06)$ & $-2.5970616(-07)$ & $+1.2274914(-08)$ & $-2.6059737(-10)$ \\
  2 &   2 &   0 &   1 & $+1.2432496(-01)$ & $-2.0329350(-04)$ & $-8.4125089(-05)$ & $-6.5746041(-06)$ & $+4.8432273(-07)$ & $-2.5618993(-08)$ & $+4.6034045(-10)$ \\
  2 &   3 &   0 &   1 & $-1.1653596(-01)$ & $+1.5588419(-04)$ & $-1.1838358(-05)$ & $+8.9824028(-06)$ & $-6.5844241(-07)$ & $+3.3233175(-08)$ & $-5.1560733(-10)$ \\
  2 &   4 &   0 &   1 & $+9.3580311(-02)$ & $-9.5460698(-05)$ & $+6.5730502(-05)$ & $-8.3498609(-06)$ & $+6.0631467(-07)$ & $-2.8836584(-08)$ & $+3.6497460(-10)$ \\
  2 &   0 &   0 &   2 & $-3.4310626(-02)$ & $-4.8748951(-05)$ & $-2.0400601(-05)$ & $-1.5151197(-06)$ & $+1.3091955(-07)$ & $-5.9460253(-09)$ & $+1.1489657(-10)$ \\
  2 &   1 &   0 &   2 & $-8.9621390(-02)$ & $+1.1162612(-04)$ & $+2.3220413(-05)$ & $+4.7241393(-06)$ & $-3.3050355(-07)$ & $+1.7133468(-08)$ & $-2.6779231(-10)$ \\
  2 &   2 &   0 &   2 & $+1.5992600(-02)$ & $+5.7868756(-05)$ & $+1.7305886(-04)$ & $-5.8547891(-06)$ & $+3.8742757(-07)$ & $-1.8796176(-08)$ & $+1.8109646(-10)$ \\
  2 &   3 &   0 &   2 & $+5.4597093(-02)$ & $-1.6763366(-04)$ & $-2.0199981(-04)$ & $+1.0071808(-06)$ & $-5.6454966(-08)$ & $-2.2273538(-09)$ & $+2.6538452(-10)$ \\
  2 &   4 &   0 &   2 & $-8.6471576(-02)$ & $+2.8303645(-04)$ & $+9.8896851(-05)$ & $+1.0678684(-05)$ & $-8.6247191(-07)$ & $+4.6701618(-08)$ & $-9.9686956(-10)$ \\
  2 &   0 &   0 &   3 & $-1.0598525(-02)$ & $-3.3242499(-05)$ & $-9.2381363(-06)$ & $-1.2554600(-06)$ & $+9.6726985(-08)$ & $-4.5448359(-09)$ & $+7.3351045(-11)$ \\
  2 &   1 &   0 &   3 & $-5.2063481(-02)$ & $+3.1653858(-06)$ & $-2.6466463(-05)$ & $+1.4009085(-06)$ & $-7.1802039(-08)$ & $+3.4452438(-09)$ & $+5.7147722(-12)$ \\
  2 &   2 &   0 &   3 & $-5.1123580(-02)$ & $+1.6569123(-04)$ & $+9.3357514(-05)$ & $+4.4470399(-06)$ & $-3.6468372(-07)$ & $+2.0844255(-08)$ & $-4.7717348(-10)$ \\
  2 &   3 &   0 &   3 & $+5.5806476(-02)$ & $-1.8013890(-04)$ & $+1.0704773(-04)$ & $-1.5950509(-05)$ & $+1.2328674(-06)$ & $-6.3734805(-08)$ & $+1.1107689(-09)$ \\
  2 &   4 &   0 &   3 & $-1.6060862(-02)$ & $+2.1128422(-04)$ & $-2.9753404(-04)$ & $+2.6744558(-05)$ & $-2.1573446(-06)$ & $+9.8821043(-08)$ & $-1.4661777(-09)$ \\
  2 &   0 &   0 &   4 & $-2.7327627(-03)$ & $-1.2167418(-05)$ & $-3.8752761(-06)$ & $-4.1236034(-07)$ & $+2.5768508(-08)$ & $-1.2520331(-09)$ & $+1.2507238(-11)$ \\
  2 &   1 &   0 &   4 & $-2.2135931(-02)$ & $-5.1249312(-05)$ & $-1.4054118(-05)$ & $-2.0999622(-06)$ & $+1.9246609(-07)$ & $-9.1132710(-09)$ & $+1.9131336(-10)$ \\
  2 &   2 &   0 &   4 & $-5.2140627(-02)$ & $+1.5426436(-04)$ & $-2.5266558(-05)$ & $+9.6175312(-06)$ & $-7.6494434(-07)$ & $+3.6497947(-08)$ & $-5.9670086(-10)$ \\
  2 &   3 &   0 &   4 & $-1.7859458(-03)$ & $-8.8721551(-05)$ & $+2.0712830(-04)$ & $-1.4925036(-05)$ & $+1.2077260(-06)$ & $-5.0800584(-08)$ & $+6.2642918(-10)$ \\
  2 &   4 &   0 &   4 & $+4.1268398(-02)$ & $+2.8661812(-05)$ & $-1.5338129(-04)$ & $+7.4383404(-06)$ & $-6.7373025(-07)$ & $+1.4045273(-08)$ & $+1.7072172(-10)$ \\
\hline
\end{tabular}

\end{table}

% include v = 1-0 A coefficients table
\begin{table}[h]
\centering
\caption{\label{tab:suppl_Acoeff} Einstein $A$ coefficients and lifetimes for the $v = 1-0$ infrared transitions of CH$^+$ ($X\,{}^1\Sigma^+$). The wavelength ($\lambda$), $m$-dependent Einstein coefficient ($A_m$), upper state energy ($E_u$, relative to the $v = J = 0$ state), upper state degeneracy ($g$), and infrared lifetime ($\tau_\text{IR}$) are tabulated for each $R$ or $P$ transition with $\vert m \vert \leq 20$. Note that $\tau_\text{IR}$ accounts for the total decay from both the $R$ and $P$ transitions of a given upper state (and does \textit{not} include pure rotational relaxation within the $v = 1$ state). }
\fontsize{8.5}{9.5}\selectfont
\begin{tabular}{rrccccc}
Transition & $m$  & $\lambda$ ($\mu$m) & $A_m$ (s$^{-1}$) & $E_u$ (cm$^{-1}$) & $g$ & $\tau_\text{IR}$ (s) \\
\hline
% LaTeX table for CH+ v = 1-0 A coefficients
% [This table was generated on Jan-21-2021 -- PBC]
% 

$R(19)$ & $ 20$ & $3.27528412$ &  $ 2.88283$  & $ 8153.6520$ & $41$ & $ 0.10688$ \\
$R(18)$ & $ 19$ & $3.27865744$ &  $ 2.41605$  & $ 7657.4062$ & $39$ & $ 0.11356$ \\
$R(17)$ & $ 18$ & $3.28357915$ &  $ 1.98722$  & $ 7182.2920$ & $37$ & $ 0.12092$ \\
$R(16)$ & $ 17$ & $3.29004857$ &  $ 1.59804$  & $ 6728.8477$ & $35$ & $ 0.12904$ \\
$R(15)$ & $ 16$ & $3.29806798$ &  $ 1.24991$  & $ 6297.5902$ & $33$ & $ 0.13797$ \\
$R(14)$ & $ 15$ & $3.30764260$ &  $ 0.94394$  & $ 5889.0140$ & $31$ & $ 0.14779$ \\
$R(13)$ & $ 14$ & $3.31878053$ &  $ 0.68091$  & $ 5503.5903$ & $29$ & $ 0.15856$ \\
$R(12)$ & $ 13$ & $3.33149263$ &  $ 0.46122$  & $ 5141.7661$ & $27$ & $ 0.17034$ \\
$R(11)$ & $ 12$ & $3.34579258$ &  $ 0.28493$  & $ 4803.9631$ & $25$ & $ 0.18317$ \\
$R(10)$ & $ 11$ & $3.36169683$ &  $ 0.15172$  & $ 4490.5771$ & $23$ & $ 0.19705$ \\
$ R(9)$ & $ 10$ & $3.37922458$ &  $ 0.06085$  & $ 4201.9770$ & $21$ & $ 0.21196$ \\
$ R(8)$ & $  9$ & $3.39839785$ &  $ 0.01117$  & $ 3938.5038$ & $19$ & $ 0.22780$ \\
$ R(7)$ & $  8$ & $3.41924147$ &  $ 0.00108$  & $ 3700.4703$ & $17$ & $ 0.24439$ \\
$ R(6)$ & $  7$ & $3.44178313$ &  $ 0.02845$  & $ 3488.1600$ & $15$ & $ 0.26146$ \\
$ R(5)$ & $  6$ & $3.46605346$ &  $ 0.09044$  & $ 3301.8266$ & $13$ & $ 0.27861$ \\
$ R(4)$ & $  5$ & $3.49208610$ &  $ 0.18326$  & $ 3141.6934$ & $11$ & $ 0.29532$ \\
$ R(3)$ & $  4$ & $3.51991780$ &  $ 0.30131$  & $ 3007.9530$ & $ 9$ & $ 0.31093$ \\
$ R(2)$ & $  3$ & $3.54958852$ &  $ 0.43506$  & $ 2900.7664$ & $ 7$ & $ 0.32473$ \\
$ R(1)$ & $  2$ & $3.58114158$ &  $ 0.56372$  & $ 2820.2628$ & $ 5$ & $ 0.33594$ \\
$ R(0)$ & $  1$ & $3.61462377$ &  $ 0.61818$  & $ 2766.5397$ & $ 3$ & $ 0.34387$ \\
$ P(1)$ & $ -1$ & $3.68758131$ &  $ 2.87367$  & $ 2739.6618$ & $ 1$ & $ 0.34799$ \\
$ P(2)$ & $ -2$ & $3.72716939$ &  $ 2.28988$  & $ 2766.5397$ & $ 3$ & $ 0.34387$ \\
$ P(3)$ & $ -3$ & $3.76891253$ &  $ 2.41302$  & $ 2820.2628$ & $ 5$ & $ 0.33594$ \\
$ P(4)$ & $ -4$ & $3.81287803$ &  $ 2.64446$  & $ 2900.7664$ & $ 7$ & $ 0.32473$ \\
$ P(5)$ & $ -5$ & $3.85913808$ &  $ 2.91483$  & $ 3007.9530$ & $ 9$ & $ 0.31093$ \\
$ P(6)$ & $ -6$ & $3.90777010$ &  $ 3.20295$  & $ 3141.6934$ & $11$ & $ 0.29532$ \\
$ P(7)$ & $ -7$ & $3.95885710$ &  $ 3.49882$  & $ 3301.8266$ & $13$ & $ 0.27861$ \\
$ P(8)$ & $ -8$ & $4.01248812$ &  $ 3.79625$  & $ 3488.1600$ & $15$ & $ 0.26146$ \\
$ P(9)$ & $ -9$ & $4.06875865$ &  $ 4.09076$  & $ 3700.4703$ & $17$ & $ 0.24439$ \\
$P(10)$ & $-10$ & $4.12777121$ &  $ 4.37871$  & $ 3938.5038$ & $19$ & $ 0.22780$ \\
$P(11)$ & $-11$ & $4.18963583$ &  $ 4.65704$  & $ 4201.9770$ & $21$ & $ 0.21196$ \\
$P(12)$ & $-12$ & $4.25447080$ &  $ 4.92310$  & $ 4490.5771$ & $23$ & $ 0.19705$ \\
$P(13)$ & $-13$ & $4.32240329$ &  $ 5.17457$  & $ 4803.9631$ & $25$ & $ 0.18317$ \\
$P(14)$ & $-14$ & $4.39357017$ &  $ 5.40944$  & $ 5141.7661$ & $27$ & $ 0.17034$ \\
$P(15)$ & $-15$ & $4.46811893$ &  $ 5.62596$  & $ 5503.5903$ & $29$ & $ 0.15856$ \\
$P(16)$ & $-16$ & $4.54620861$ &  $ 5.82263$  & $ 5889.0140$ & $31$ & $ 0.14779$ \\
$P(17)$ & $-17$ & $4.62801096$ &  $ 5.99824$  & $ 6297.5902$ & $33$ & $ 0.13797$ \\
$P(18)$ & $-18$ & $4.71371164$ &  $ 6.15178$  & $ 6728.8477$ & $35$ & $ 0.12904$ \\
$P(19)$ & $-19$ & $4.80351168$ &  $ 6.28250$  & $ 7182.2920$ & $37$ & $ 0.12092$ \\
$P(20)$ & $-20$ & $4.89762896$ &  $ 6.38985$  & $ 7657.4062$ & $39$ & $ 0.11356$ \\
\hline
\end{tabular}

\end{table}

\end{document}